\documentclass{kapproc} 
\kluwerbib

\begin{document}

\articletitle[Time Lags of Z Source GX~5-1]
{Time Lags of Z Source GX~5-1}

\author{J. L. Qu$^{(1)}$, Y. Chen$^{(1)}$, M. Wu$^{(1)}$, L.
Chen$^{(1,2)}$, L. M. Song$^{(1)}$}
\affil{$^{(1)}$Laboratory for Particle Astrophysics, Institute of High Energy
Physics, CAS, Beijing 100039, P. R. China\\
$^{(2)}$Department of Astronomy, Beijing Normal University, Beijing
100875, P. R. China}
\email{qujl@mail.ihep.ac.cn}


\begin{abstract}
We investigated the time lags and the evolution of the cross spectra
of Z source GX~5-1, observed by the Rossi X-ray Timing Explorer
(RXTE), when it is in the horizontal branch oscillations. We showed
that the time lags of 3 horizontal branch oscillations are related to
the position on the hardness intensity diagram. All of the three QPOs
were shown to have hard time lags. However on the cross spectra, one
is in a `dip', one in a `bump', the other has no so obvious
characteristic. The time lags of two of the QPOs decrease
with QPO's frequency, while the other has a trend increasing
with its frequency. Moreover, in the normal branch, we found no
significant time lags in the present observational data.
\end{abstract}


\section{Introduction}
The neutron-star low-mass X-ray binaries (LMXBs) can be divided into
two types on the basis of pattern traced out in the X-ray
intensity-hardness or color-color diagram (HID or CCD, respectively;
Hasinger \& van der Klis 1989). One is Z source which traced out a
``Z" pattern on HID or CCD, the other is atoll source which traced out
a ``C" pattern on CCD. From the top to the bottom of the Z pattern,
the three limbs of the pattern traced out on HID or CCD is called the
Horizontal Branch (HB), Normal Branch (NB), and the Flaring Branch
(FB), respectively. The temporal properties of Z source depend on the
position on the Z pattern. On the HB and the NB, there are varying
quasi-periodic oscillations (QPOs) with frequency less than 100 Hz,
called horizontal branch oscillation (HBO) and normal branch
oscillation (NBO).

Recently, Jonker et al. (2002) studied in detail the power-density
spectra (PDS) of the Z source GX~5-1. They showed that there are four
QPO components on the HB, which are HBO and its three harmonics.
Using the EXOSAT data, van der Klis et al. (1987) had studied
the time lags (or phase lags) of the HBO and the low frequency noise
(LFN) by cross-correlation spectra, showing a complex timing behavior.
There are hard time lags decreasing with the HBO frequency and soft
time lags in LFN increasing with the Fourier frequency. They argued
that the hard time lags of HBO are due to Comptonization of soft
photons, and the soft time lags of LFN due to the evolution of
energy spectrum of the shot. Using the Ginga data, Vaughan et al.
(1994) studied one harmonic with higher Fourier frequency and the
relation between time lags of NBO and its energy, in which a jump
at 3.5 keV was found. Their results are similar to that from van der
Klis et al. (1987).

Thanks to the large effective collective area and high time-resolution
of the Rossi X-ray Timing Explorer (RXTE), we can investigate the
cross-spectrum and PDS of GX~5-1 in more details. In this paper, we
report the time lags of HBO and its harmonics of GX~5-1 basing on the
RXTE data. We found that the time lag behaviors are related
to the HBO fundamental frequency. In section 2, we describe the method
of data analysis. In section 3, the results are presented. We discuss
the results briefly in the last section.

\section{Observations and Analysis}

GX~5-1 was observed for 11 times in 1997 with the Proportional Counter
Array (PCA) on board the RXTE satellite. Because the energy channel
bins changed during the different observation epochs, we only used
data of 6 observations from May 30 to July 25, 1997. The data were
always obtained in a mode with 16 s time-resolution and a high
spectral resolution (129 energy channels covering the effective 2-60
keV range, the Standard 2 mode). For this epoch of observations we
used Single-Bit mode, a high time-resolution data mode with $2^{-13}$
s time-resolution covering four energy bands (namely 2-5.1
keV, 5.1-6.6 keV, 6.6-8.7 keV, and 8.7-60 keV; absolutely RXTE energy
channels 0-13, 14-17, 18-23, 24-249, respectively), to calculate PDS
and cross spectrum.

We used standard 2 mode data to make an HID for GX~5-1. Because one
detector was off sometimes, we only used data when all 5 detectors
were on simultaneously. The hardness, or hard color, is defined as the
8.7-19.7keV/5.9-8.7keV (24-53/l6-23) count rate ratio and the
intensity is defined as the count rate in the 2-19.7 keV (0-53) band
averaged in a 32 s.

The HID was showed in Fig.1, each point represents 32 s data length in
the HID. The source was found on the HB and the NB. To investigate
X-ray temporal properties of GX~5-1
along the Z track, we divided the HB into 6 boxes (see Fig.1),
according to the point numbers of each box.

\begin{figure}[ht]
\caption{The HID of GX~5-1. Each point corresponds to 32 s of data.}
\end{figure}
\vskip 0.1cm

The good quality of data with high time resolution allows us to
calculate the power-density spectra
for each energy band and the cross-spectra for two energy bands of data
stretches of 32 s with 4 ms time bin in each box. All the spectra were
logarithmically re-binned. The cross spectra of GX~5-1 show that time
lags above 80 Hz are dominated by the dead-time effect, which
was corrected by subtracting a cross vector, averaged over 80 to 128
Hz, from the cross spectrum (van der Klis et al. 1987; Qu, Yu \& Li
2001). As an illustration, only the PDSs and the cross spectra of the
highest energy band are shown in Fig.2

\begin{figure}[ht]
\caption{The PDSs for 8.7 -- 60 keV ({\emph{top panel}}) and cross spectra
between 2 -- 5.1 keV and 8.7 -- 60 keV ({\emph{bottom panel}}) of GX~5-1 on
different boxes.}
\end{figure}

\begin{table*}[t]
\caption[]{The centroid frequencies and the Full Width at Half
Maximum of 3 QPOs.}
\label{Table 1}
\begin{tabular}{crrrrrc}
\hline\hline
Box No.& sub-HBO (Hz) & FWHM (Hz)  & HBO (Hz) &   FWHM (Hz) & harmonic (Hz) & FWHM (Hz)\\ \hline
1&8.15$\pm$1.20 &8.17$\pm$3.10&17.39$\pm$0.09&3.74$\pm$0.28&31.34$\pm$1.19&8.17$\pm$2.63\\
2&10.29$\pm$1.13&8.78$\pm$3.78&22.29$\pm$0.22&5.81$\pm$0.47&42.83$\pm$1.41&20.69$\pm$2.67\\
3&11.96$\pm$1.37&12.30$\pm$3.21&26.69$\pm$0.11&5.04$\pm$0.50&50.85$\pm$1.28&22.51$\pm$3.13\\
4&17.64$\pm$2.62&11.58$\pm$12.28&32.37$\pm$0.18&7.15$\pm$0.57&62.11$\pm$1.50&19.16$\pm$4.38\\
5&19.68$\pm$1.65&2.31$\pm$18.87&37.29$\pm$0.32&8.62$\pm$1.07&72.35$\pm$3.78&19.38$\pm$11.03\\
6&22.07$\pm$5.19&2.86$\pm$---&41.41$\pm$0.50&10.52$\pm$2.01&86.31$\pm$8.03&28.59$\pm$34.52\\ \hline
\end{tabular}
\end{table*}

\section{Results}

Figure 2 ({\it panel 1--6, top}) presents that there exist two obvious
QPOs, and their frequencies increase from `Box l' to `Box 6'. Jonker
et al. (2002) showed that the PDS of GX~5-1 can be fitted well by a
model consisting of 4 QPO components (Lorentzians) and a cut-off
power-law component. We only used a model composed of three
Lorentzians (i.e., 3 QPO components) and a cut-off power-law to fit
the PDSs of GX~5-1, while the harmonic at high frequency has not been
studied due to the low signal-to-noise ratio caused by the limited
integration time. The QPO parameters were listed in table 1. Following
Jonker et al. (2002), we call the low-frequency QPO as sub-HBO, and
the strongest QPO as HBO, the higher frequency QPO as the harmonic. In
Fig.2, QPOs' position (QPO's centroid frequency) were indicated by the
dotted-lines. As an illustration, we plotted the HBO vs. sub-HBO and
the harmonic in Fig.3. The results are consistent with those of Jonker
et al. (2002).

The cross spectra (Fig.2; {\it panel 1--6, bottom}) show that the time
lags of the noise less than l0 Hz are dominated by soft time lags
(negative lags are represented by `$\circ$' in the cross spectra of
Fig.2). During QPOs' frequency range, GX~5-l shows hard time lags
(positive lags are presented by `+' in the cross spectra of Fig.2). 
The time lags of HBO and the harmonic decrease with their centroid 
frequencies(see Fig.4), but
the time lag of HBO in box 6 has a trend increasing with QPO frequency.
Those results are consistent with earlier works (van der Klis et al.
1987, Vaughan et al. 1994). However, the cross spectra of GX~5-1 also
show finer timing behaviors. The time lags of the harmonic and the HBO
may be in a bulge and a concave, which were named as `bump' and `dip'
respectively in this paper, for convenience. For checking those
characteristics in the cross spectra, we fitted the cross spectrum
with a quadratic and a Lorentzian around the centroid frequency of the
harmonic and the HBO. We found that the centroid frequency of the HBO
and the harmonic is linear with the fitted results of the cross
spectra, with a slope near 1.0. We also calculated the average time
lags of the HBO and the harmonic. In the first five boxes it is found
that the time lag of the harmonic is higher than the HBO's with a
significance level $\sim 4.1 \sigma$, which is obtained from numerical
simulations. Those results showed that the time lags of the HBO and
the harmonic are in a dip and a bump respectively. Due to the
contamination of the LFN, the behavior of the sub-HBO may  also be
different from the those of the HBO and the harmonics. The time lags
of the sub-HBO has a trend increasing with its frequency(see bottom
panel of Fig.5). We also calculated the cross spectrum of `Box NB' on
the NB and average the time lags in NBO frequency range. However, we
didn't find any significant time lags at 90\% confidence level
(average time lag of the NBO is 4.9$\pm$8.0 ms).

In order to study the relations between the time lags of different
QPOs, we average the time lags over the frequency ranges from
$\nu_{\rm QPO}-0.5\Delta\nu_{\rm QPO}$ to $\nu_{\rm QPO}+
0.5\Delta\nu_{\rm QPO}$,  where $\Delta\nu_{\rm QPO}$ is the full
width at half maximum (FWHM) of the QPO. The time lags of the HBO and
the harmonic were plotted in Fig.4. The time lags of the sub-HBO were
plotted in Fig.5(bottom panel). Because the sub-HBO was in the
frequency range of the LFN, time lags of the LFN might have
affected the sub-HBO. We fitted the real and imaginary parts of a
cross vector between 1 and 5 Hz by Lorentzian and quadratic model
respectively. Then the time lags of the LFN were corrected by
subtracting a real and imaginary part from the cross spectra of the
sub-HBO frequency range basing on the above models. The results were
shown in top panel of the Fig.5.

\begin{figure}[ht]
\sidebyside
{\caption{The relations among HBO and
sub-HBO ({\emph bottom panel}) as well as the harmonic ({\emph top
panel}).}}
{\caption{Time lags of HBO and the harmonic.}}
\end{figure}
\vskip 0.1cm

In order to discuss the thermal electron distribution near the neutron
star, we used all the data from Box1$\sim$3 and calculated the
coherence function which shows how the photons in the two energy bands
are related. The results were shown in Fig.6. For clear, the coherence
functions for high energy bands are offset by 1.0 and 2.0 respectively.

\begin{figure}[ht]
\sidebyside
{\caption{Time lags of sub-HBO. The top is
the time lags corrected with LFN's by subtracting a real and imaginary
part from the cross spectra(see text).}}
{\caption{The coherence function of GX~5-1 on the HB. The upper two
panels are offset by 1.0 and 2.0 respectively.}}
\end{figure}
\vskip 0.1cm

For studying the energy dependence of the lags, the cross
spectra between higher energy bands and 0-13 energy band was calculated
in each box. The energy dependence of the HBOs was listed table 2.
For simple, only the energy dependence of time lags of 17 Hz QPO
were shown in Fig.7.
\begin{table*}[t]
\caption[]{The energy dependence of time lags of HBOs.}
\label{Table 2}
\begin{tabular}{crrrrc}
\hline\hline
Box No.& HBO&FWHM & $E_2=5.78^{0.80}_{-0.64}$&$E_3=7.50^{+1.24}_{-0.92}$&$E_4=11.1^{+49.0}_{-2.36}$\\ \hline
1 &17.39&3.74&   0.53$\pm$0.11     &1.01 $\pm$0.11 &1.58$\pm$0.12\\
2 &22.29&5.81&   0.29$\pm$0.12     &0.74 $\pm$0.12 &1.15$\pm$0.12\\
3 &26.69&5.04&   0.19$\pm$0.09     &0.36 $\pm$0.08 &0.83$\pm$0.09\\
4 &32.37&7.15&   0.00$\pm$0.10     &0.17 $\pm$0.10 &0.66$\pm$0.09\\
5 &37.29&8.62&  -0.27$\pm$0.19     &0.15 $\pm$0.16 &0.47$\pm$0.13\\
6 &41.41&10.5&   0.12$\pm$0.28     &0.12 $\pm$0.24 &0.49$\pm$0.21\\
NB& 5.69&4.0 &   1.76$\pm$7.94     &0.82 $\pm$9.38 &4.90$\pm$7.99\\ \hline
\end{tabular}
$^*$The unit of time lag and energy bands is ms and keV respectively.\\
$E_1=3.93_{-3.82}^{+1.21}$(keV)
\end{table*}

\begin{figure}[ht]
{\caption{The time lags of 17 Hz QPO as a function of energy compared to the
lowest energy 2-5 keV(channel 0-13).}}
\end{figure}
\vskip 0.1cm

\section{Discussion}
We have performed the analyzed of the RXTE/PCA data of GX~5-1 when it
was on the HB and the NB. Beside that the soft time lags of the LFN
and the hard time lags of QPOs were found to be similar to the results
of Vaughan et al. (1994), we discovered some finer characteristics in
the cross spectra around the centroid frequency of the QPOs. The time
lag of the HBO is in a dip while the harmonic in a bump. The sub-HBO
also shows a hard time lag. Due to the contamination of the LFN, the
time lag of the sub-HBO increases with the centroid frequency of the
sub-HBO. When the effect of the LFN was corrected, this trend
disappear. The QPOs in the horizontal branch show hard time lag. In
present RXTE data and energy bands, we didn't find any significant
time lags in the NBO's frequency range. It may be caused by that the
lowest energy channel for calculating cross spectra is above the jump
energy (3.5 keV) of time lags (Vaughan et al. 1999).

If box series numbers in the HID represent the mass accretion rate of
the source, the evolution of the cross spectrum along the track
suggests that the cross spectrum should vary with the mass accretion
rate.

There are several kinds of models proposed to interpret the time lags:
1) the Comptonization models, e.g., the uniform corona model (Payne
1980), 2) the non-uniform corona model (Kazanas, Hua, \& Titarchuk
1997), and 3) the drifting-blob model (B\"ottcher \& Liang 1999), they
only explain the hard time lags and the energy dependence of the time lags.
The soft time lags cannot be
explained. However, those models have some problems physically. For
example, for producing measured time lags in a static Compton cloud, a
hot corona with a large radius (for example, for Cyg~X-1, the radius
extent of the hot corona exceeds $10^{10} r_g$), it is physically
unrealistic (see the review by Poutanen 2000). The coherence function
might reflect the dynamical properties of the corona in the corona
models(Nowak et al. 1999a, 1999b, Ji et al. 2003). The loss of
coherence in high energy channels might show that
the corona of GX~5-1 is dynamical. This is consistent with the above
suggestions.

To explain the soft time lags and the time lag evolution in
GRS~1915+105 and neutron-star binary Cir~X-1 (Cui 1999; Reig et al.
2000, Qu, Yu \& Li 2001), a two-layer corona model has been proposed
(Nobili et al. 2001, Qu, Yu \& Li 2001). In this model, the evolution
of time lags of GX~5-1 with mass accretion can be explained, but the
increase of the time lag near the transition between the HB and the NB
can not be explained.

Both shot profile properties and Comptonization of photons can
introduce time lags. The numerical simulations showed that the
energy-dependent shot profiles can produce low-energy phase lags in
the cross spectrum at typical frequencies for the shot time scale
(tenths of a Hz to a few Hz) without noticeably affecting the
cross spectrum at higher frequencies(Shibazaki et al. 1988). The shots
are thought to originate near or at the neutron star surface as
material falls through the magnetosphere and onto the surface of the
neutron star. The delays should be of the order of the free-fall time.
But the free-fall time is not longer than 0.5 ms. Although
the shot models can produce almost any time-delay spectrum like
Comptonization models, they also have problems on physical grounds
(see Vaughan 1994, Poutanen 2000, Qu et al. 2002).

\begin{acknowledgments}
We would like to thank the anonymous referee for carefully reading
this manuscript and for helpful comments. J.L. Qu also thanks J.-M.
Wang for his reading of the
manuscript. We are grateful to R. Staubert for help in downloaded the
data. This work was subsidized by the Special Funds for Major State
Basic Research Projects and by the National Natural Science Foundation
of China.
\end{acknowledgments}

\begin{figure}[ht]
\vskip 2in
\includegraphics{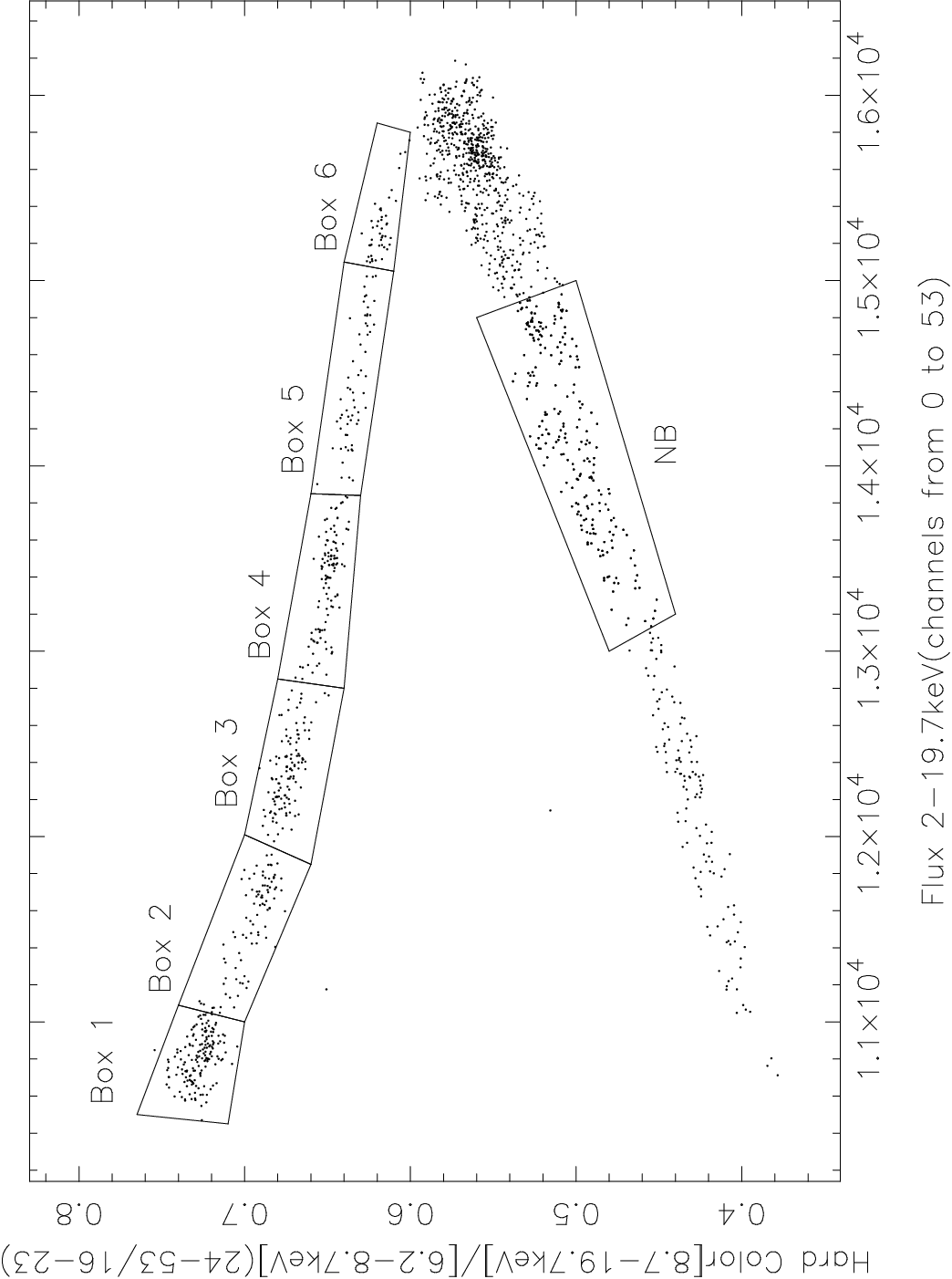}
\vskip 4in
\end{figure}

\begin{figure}
\vskip 2in
\includegraphics{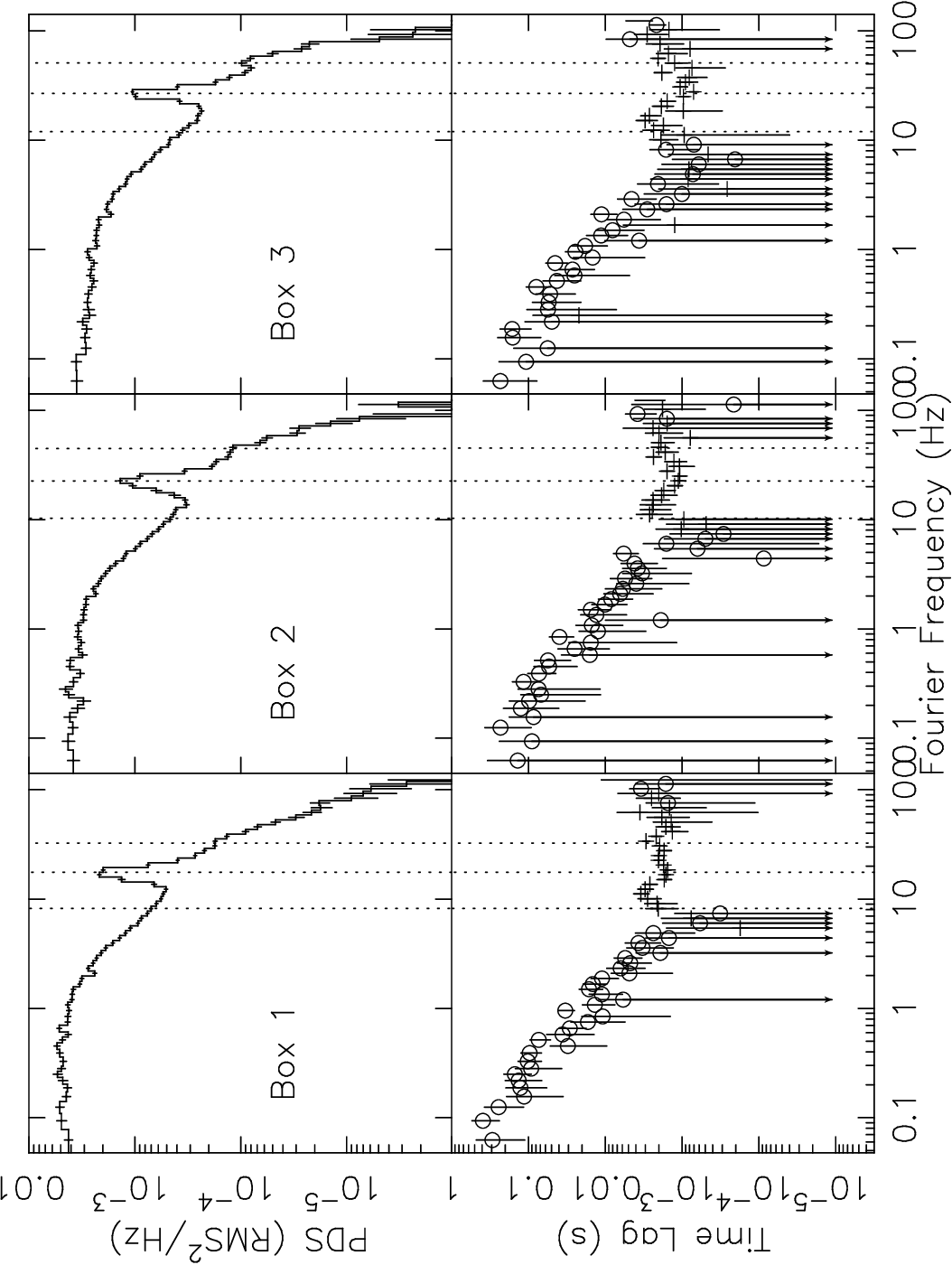}
\end{figure}
\newpage

\begin{figure}
\vskip 4in
\includegraphics{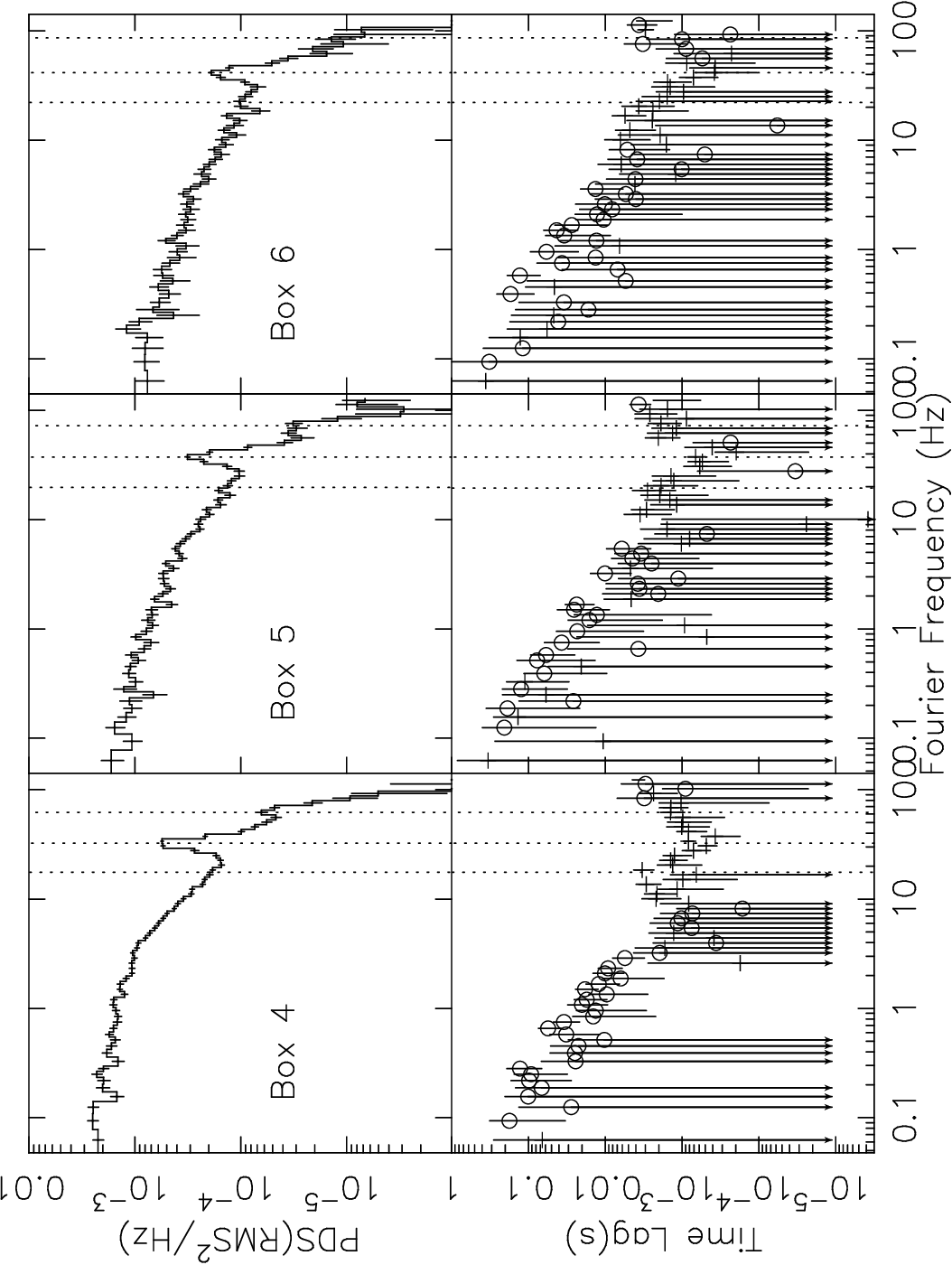}
\vskip 2in
\end{figure}
\newpage

\newpage
\begin{figure}
\vskip 2in
\includegraphics{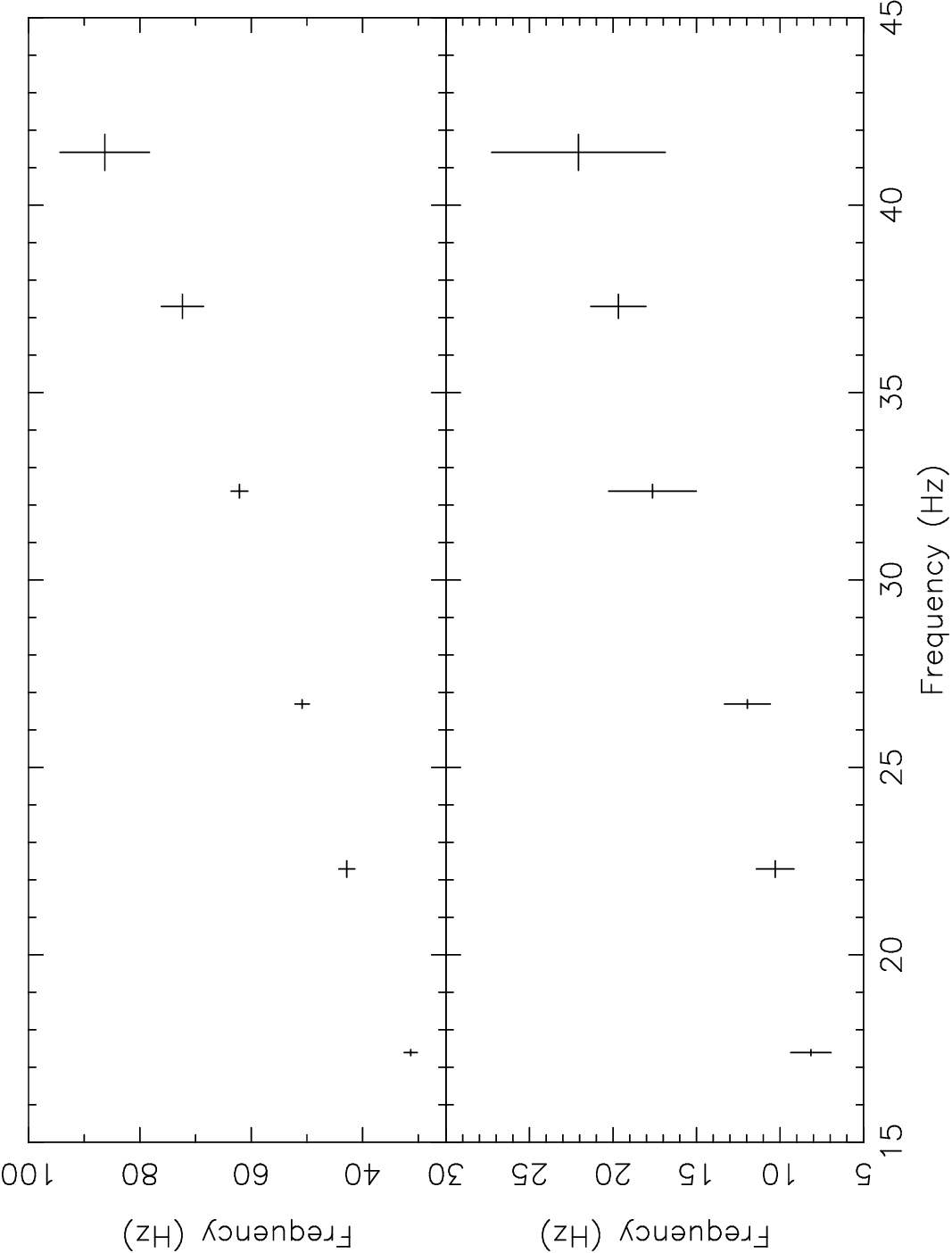}
\end{figure}
\newpage

\newpage
\begin{figure}
\vskip 4in
\includegraphics{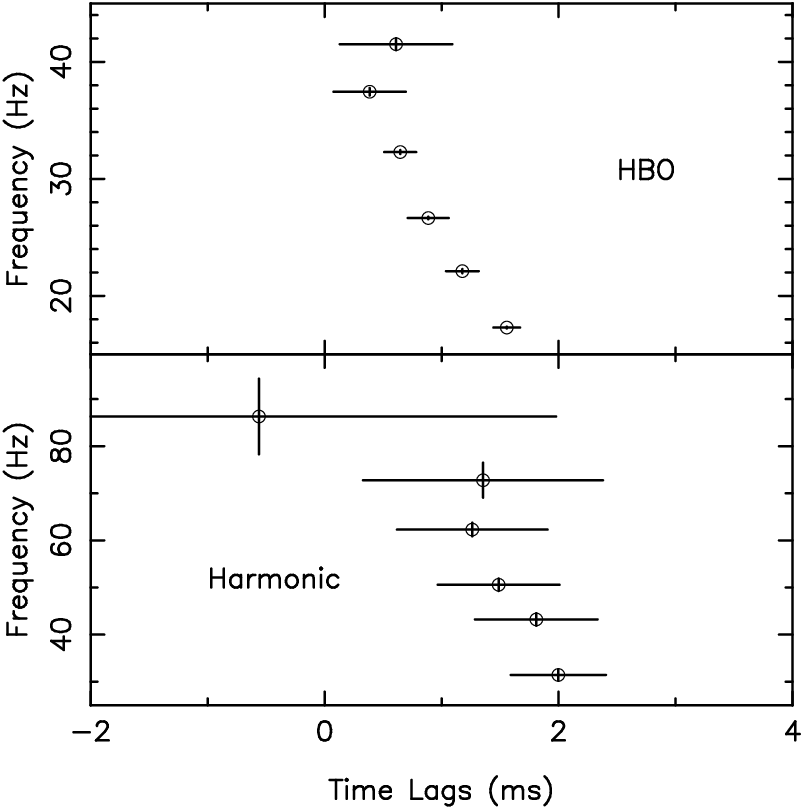}
\end{figure}
\newpage

\newpage
\begin{figure}
\vskip 4in
\includegraphics{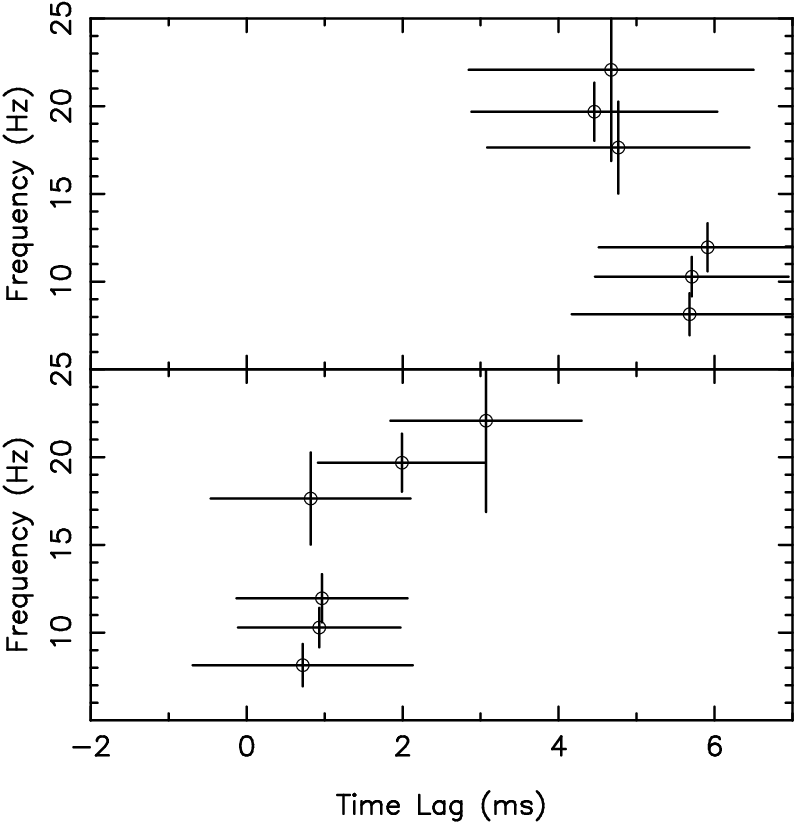}
\end{figure}
\newpage

\newpage
\begin{figure}
\vskip 4in
\includegraphics{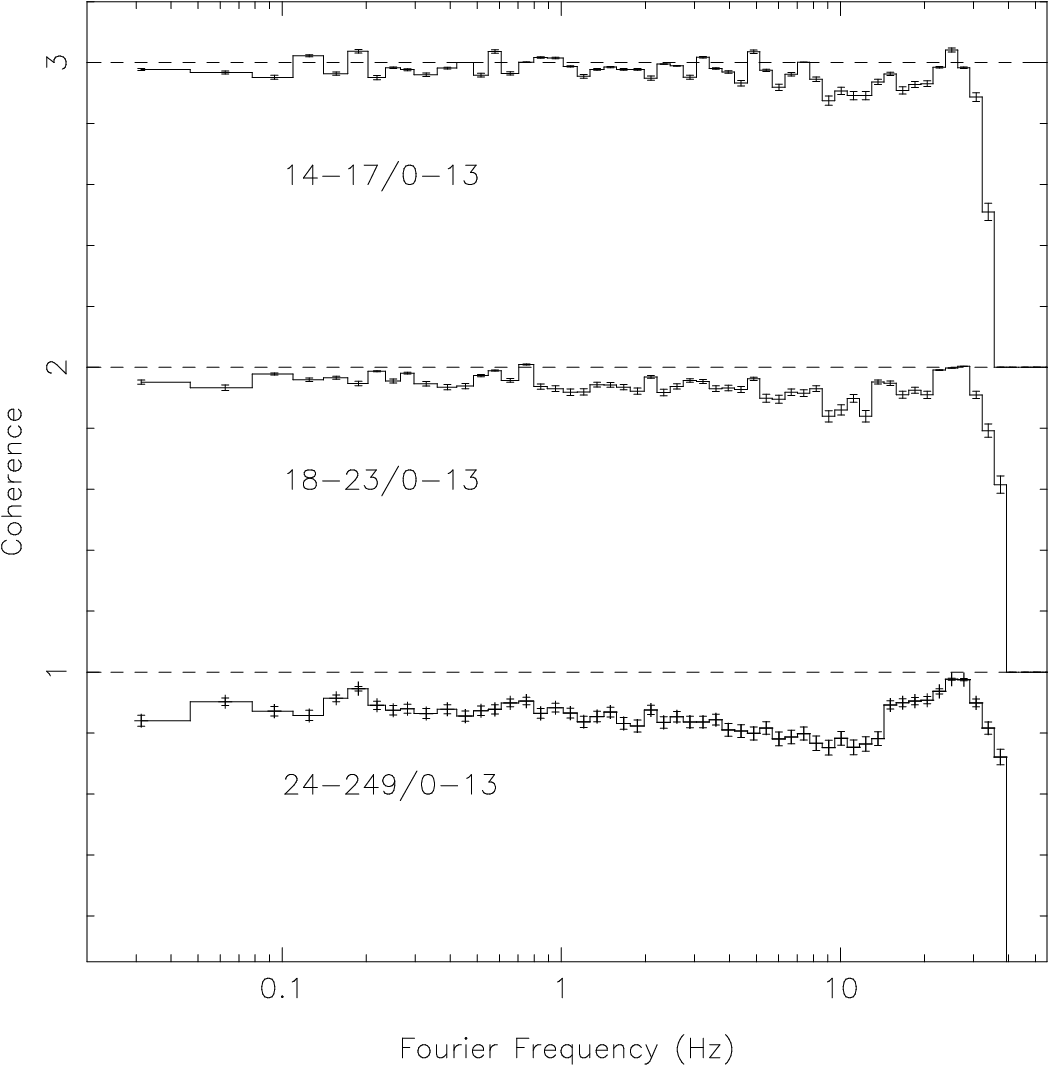}
\end{figure}
\newpage

\newpage
\begin{figure}
\vskip 4in
\includegraphics{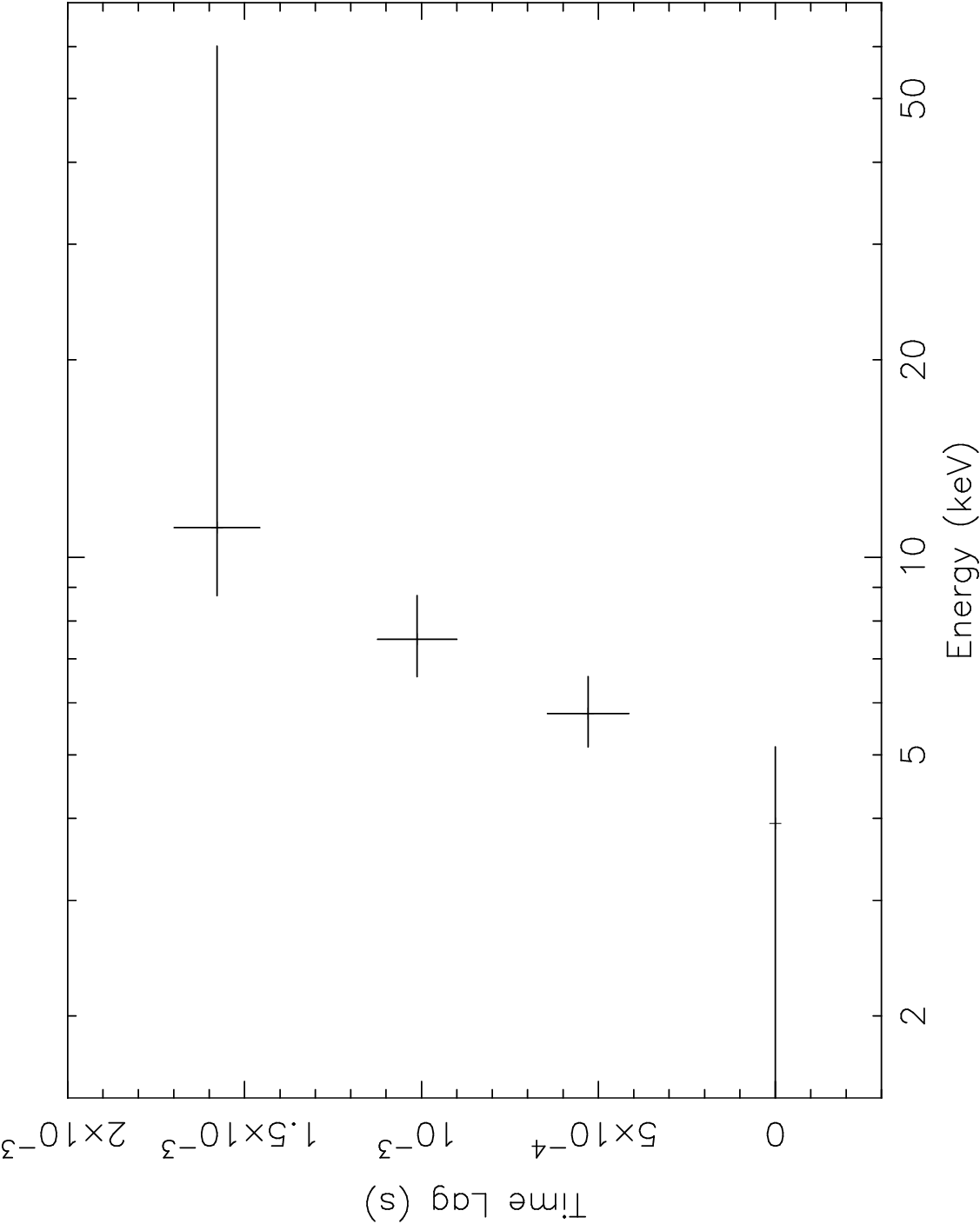}
\end{figure}
\newpage


\end{document}